\documentclass[aps,preprint]{revtex4}

\usepackage{graphicx}

\begin{document}

\date{\today}

\title{Non-Hermitian Quantum Physics of Open Systems} 

\author{
Ingrid Rotter\footnote{email: rotter@pks.mpg.de}}
\affiliation{
Max Planck Institute for the Physics of Complex Systems,
D-01187 Dresden, Germany  }

\vspace*{1.5cm}

\begin{abstract}

Information on quantum systems can be obtained only when they are
open (or opened) in relation to a certain environment.  
As a matter of fact, realistic open quantum systems appear in 
very different shape. We sketch the theoretical description
of open quantum systems by means of a
projection operator formalism  elaborated many years ago, 
and applied by now to the description of different open
quantum systems. The Hamiltonian describing the open quantum system is
non-Hermitian. 
Most studied are the eigenvalues of the non-Hermitian Hamiltonian of  
many-particle systems embedded in
one environment. We point to the unsolved problems of this method when
applied to the description of realistic many-body systems.   We then
underline  the role played by the
eigenfunctions of the non-Hermitian Hamiltonian. 
Very interesting results originate from the fluctuations of
the eigenfunctions in systems with gain and loss
of excitons. They occur with an
efficiency of nearly 100\%. An example is the photosynthesis.

\end{abstract}

\maketitle

1. Introduction.

What have, for example, nuclear physics and photosynthesis in common?
The quick answer is:
nothing. That is however only half the truth:
both are  open quantum systems.
The atomic nuclei are embedded in the continuum of scattering
wavefunctions whereby their states get, generally, a finite lifetime.
The photosynthesis occurs in 
the light-harvesting complex, in which light energy is converted in
chemical energy.
These two  physical processes are completely different from one another. 
The only common feature is that the two systems have a finite extension and 
are embedded in an infinitely extended environment. 
In the following, we will consider similarities and differences
between these two (and many other) so unequal open quantum systems.
\cite{comment1}. 

\vspace{.6cm}

2. Projection operator formalism.

A convenient method to describe the
different states of  a nucleus is to use the projection operator
formalism  elaborated many years ago
\cite{feshbach}, according to which
the whole function space is divided into the two subspaces $\cal P$
and $\cal Q$ with ${\cal P} + {\cal Q} = 1$. The subspace $\cal Q$
describes the finite system (e.g. the nucleus), while the subspace 
$\cal P$ contains the continuum of scattering wavefunctions.
The solution for the nucleus can be found. 
According to Feshbach, it is, however, a formal solution since
the number of states of a heavy nucleus is very large and  does not
allow to obtain numerical results without introducing approximations.
Mostly, statistical methods are used. This resulted 
finally not only in the description of nuclear states, but
in numerous studies on quantum chaos (see e.g. \cite{stockmann}).  

The situation is different in light nuclei where the level density
is small and the properties
of individual nuclear states can be investigated. Here, 
the solution obtained for the description of the nucleus is no
longer a formal solution. 
It can rather be found numerically. It is 
the basic equation of the so-called continuum shell model that
describes  the open nuclear system.  
The Hamiltonian of the continuum shell model is non-Hermitian.

Practical problems in finding reliable solutions for light nuclei in
the framework of the continuum shell model
arise from the fact that the  interaction between the nucleons in
nuclei is different from that between free nucleons. It is a residual
interaction which remains after the central mean field of the nucleus  
has been created  by the nucleons themselves. The residual 
interaction is badly known.   

These problems do not occur when dealing with atoms. In this case, 
the electrons move in the field 
that is created mainly by the atomic nucleus.
Thus, the interaction between the electrons is quite well known. Here,
the continuum shell model has been applied successfully to the
description of different interference processes, 
for references see the review \cite{top}.

\vspace{.6cm}

3. Effective non-Hermitian Hamilton operator.

Many numerical calculations for open many-body quantum systems are
performed successfully
by using the non-Hermitian Hamilton operator \cite{top} 
$$H^{\rm eff} = H_0 + 
\sum_c V_{0c}\frac{1}{E^+ - H_c}V_{c0} \; .
$$
Here, $H_0$ is the Hamiltonian describing the 
corresponding closed system with discrete states,
$(E^+ - H_c)^{-1}$ is the Green function in the 
continuum with the Hamiltonian $H_c$ describing the environment of
decay channels, and $V_{0c}, ~V_{c0}$ stand for the coupling of the
closed system to the different channels $c$ of the environment.
The non-Hermiticity of $H^{\rm eff}$ arises from the
second term of $H^{\rm eff}$, i.e. from the
perturbation of the  system occurring under the 
influence of its coupling  to the environment. 
It is complex: the real part arises from the principal value integral
and the imaginary one from the residuum (for details see \cite{top}).

Sometimes, the effective Hamiltonian is approximated by
$H^{\rm eff} = H_0 + \alpha  W $ 
where $W$ is imaginary, and the properties of the
system are studied as a function of $\alpha$ (for examples see the
review \cite{top}). In \cite{sergi}  the non-Hermitian
operator is assumed to be $\hat {\cal H} = \hat H - i \hat \Gamma$. 
This operator allows us to study the general meaning of the imaginary
part of the non-Hermitian Hamiltonian. The information on the
considered physical system, involved in  $H^{\rm eff}$, is however lost.
In any case, the non-Hermiticity
of $H^{\rm eff}$ arises from the second term which is added  
to $H_0$ as a perturbation.

Some experimentally well-known results can be understood 
when the system is described by the Hamiltonian $H^{\rm eff}$.
An example is the variation of the cross section picture at 
low level density to that at high level density. At low level density 
isolated resonances can be identified, while this is impossible
at high level density where the individual resonances overlap  
and cause a fluctuation picture of the cross section. This transition
has been traced
successfully in nuclear physics, many years ago, by means of 
$H^{\rm eff}$ \cite{ro91}.
Further examples from recent studies on mesoscopic 
and other systems can be found in the review \cite{ropp}.

The advantage of using $H^{\rm eff}$ is, above all, that the 
results for the closed system obtained with the Hermitian Hamiltonian
$H_0$,  can be used.
With vanishing coupling strength between system and environment,
the second term of $H^{\rm eff}$ vanishes, i.e. 
$H^{\rm eff}  \to H_0$.
    
In the non-Hermitian formalism there are singular points.
From mathematical studies, their existence has been known 
for quite a long time  \cite{kato}.
At these points, called mostly {\it exceptional points (EPs)}, 
two eigenvalues 
of the non-Hermitian operator coalesce causing interesting topological
effects. These effects are interesting in themselves, and are
studied not only theoretically but also experimentally. 
The influence of EPs onto the dynamics of open quantum systems is
studied in many papers, see e.g. the reviews \cite{mois,top,ropp}
and the recent papers \cite{ep}.

As a result of the many studies in different fields of physics, 
the following can be stated.
Far from EPs, Hermitian quantum physics provides usually good results.
Under the influence of EPs, however, interesting new features 
appear (which are mostly counterintuitive). 

Some problems with  $H^{\rm eff}$ are the following.\\
(i) According to numerical studies, EPs do not play any role when
the system is coupled to only one well-defined environment (called
usually  {\it one-channel case}). This  is due to
nonlinear effects involved in the 
non-Hermitian Hamiltonian \cite{pra95}. This result
agrees with the experience obtained from the successful description of
one-channel processes without taking into account EPs.\\
(ii) EPs are defined mathematically for systems coupled to one channel
\cite{kato}, while most
physically interesting processes (such as transfer processes)
occur when the system is coupled to at least two channels.

\vspace{.6cm}

4. Eigenfunctions of the non-Hermitian operator.

In order to study the influence of EPs on the properties of physical
systems one has, first of all, to know the eigenvalues of the
non-Hermitian Hamiltonian. Accordingly, the
properties of the eigenfunctions  of
$H^{\rm eff}$ are  seldom studied although they
influence the system properties not only in a larger parameter range
than the eigenvalues. Much more important (and almost unknown
in current literature) is that the eigenfunctions 
incorporate more sensitive signatures of the EPs than the eigenvalues.

The eigenfunctions $\Phi_i$ of the non-Hermitian Hamiltonian 
are, generally,  biorthogonal and should be normalized according to
\cite{top}
$$\langle \Phi_i^*|\Phi_j\rangle = \delta_{ij} \; $$
in order to guarantee a smooth transition from an open 
quantum system to a nearly closed one. 
Further, the  $\Phi_i$ are (almost) orthogonal far from EPs and differ
from one another only by a phase in approaching an EP.
Thus, the phases are not rigid, generally. The phase rigidity defined
by \cite{top}
$$
r_k ~\equiv ~\frac{\langle \Phi_k^* | \Phi_k \rangle}{\langle \Phi_k 
	| \Phi_k \rangle}  $$
characterizes the distance of the state $k$ from an EP:
$r_k \to 0$ in approaching an EP while $r_k \approx 1$ far from an EP.
Since, moreover, every state of the system is coupled to the
common environment, all states may mix via the environment.
This so-called external mixing of the eigenfunctions $\Phi_i$ is a
second-order effect which becomes infinitely large at an EP
\cite{top}.
  
The eigenfunctions $\Phi_i$ are therefore not only different from the
eigenfunctions $\Phi_i^0$
of the Hermitian Hamiltonian $H_0$. They contain, moreover, 
valuable information on the 
possible influence of an EP on the dynamics of an open quantum system
which is determined by the
distance of the considered eigenstates $i$ from an EP.
This information is involved, above all, in the phase rigidity
and in the external mixing of the eigenfunctions 
$   \Phi_i$ of the non-Hermitian Hamiltonian.

One of many examples on the role of the eigenfunctions of the
non-Hermitian Hamiltonian in physics is that higher-order EPs 
(at which more than two eigenvalues coalesce) cannot
be observed. They can be found mathematically by considering the 
eigenvalues.  However, the 
external mixing of the eigenfunctions via the environment 
causes some energy shift of all the eigenstates. As a consequence,
the possibility to 
observe higher-order EPs is prevented \cite{pra95}.

\vspace{.6cm}

5. Genuine non-Hermitian Hamilton operator.

The description of the properties of open quantum systems by means of
$H^{\rm eff}$ has provided very many interesting results which
contribute to a better understanding of
non-Hermitian many-body quantum physics.
Effects which are caused by interferences between the eigenfunctions 
$\Phi_i$ of the non-Hermitian Hamiltonian 
are however more difficult to extract or are
not at all describable with $H^{\rm eff}$.

As has been shown in \cite{muro}, interferences between the
eigenfunctions $\Phi_i$ of the non-Hermitian Hamilton operator
may become important in physical processes. For example, they
explain the so-called phase lapses observed experimentally a 
few years ago in the transmission
through quantum dots \cite{lapses}. These experimental results
remained puzzling for many years.
 
Furthermore, 
the results of non-Hermitian quantum physics are always different 
from those of Hermitian quantum physics as
has been shown theoretically as well as experimentally \cite{savin}.
This is incompatible with $H^{\rm eff}  \to H_0$ 
occurring with vanishing coupling strength between system and 
environment. It means rather that the
non-Hermitian Hamiltonian $\cal H$ must be a genuine non-Hermitian  
Hamiltonian (instead of  $H^{\rm eff}$).

A genuine non-Hermitian operator involves the complex energies 
$\varepsilon_i = e_i + i/2~\gamma_i$ \cite{comment2} 
of all $N$ states 
of the system. For $N=2$, it reads 
$$
{\cal H} = 
\left( \begin{array}{cc}
	\varepsilon_{1} \equiv e_1 + \frac{i}{2} \gamma_1  
	& ~~~~\omega   \\
	\omega 
	& ~~~~\varepsilon_{2} \equiv e_2 + \frac{i}{2} \gamma_2   \\
\end{array} \right) \; . $$
The operator $\cal H$ allows us to describe
not only quantum systems with excitation of eigenstates 
but also those  
which occur without excitation of any eigenstates of 
$\cal H$. In the last case, fluctuations of the eigenfunctions cause 
measurable effects. The fluctuations are large, above all, near to EPs.
Since  eigenstates of $\cal H$ are not excited in such a case, the
process occurs with an efficiency of nearly 100 \% \cite{pre95}. 

This theoretical result allows us to obtain an understanding for the
experimental results obtained for the  photosynthesis 
\cite{engel,engel-ed,fleming,romero}. 
These experimental results  cannot be described by means of the
standard Hermitian operator $H_0$ or by using 
the effective Hamiltonian  $H^{\rm eff}$ which
both describe the many-particle properties of the system with
excitation of eigenstates of the Hamiltonian.
Fluctuations being basic in the process of photosynthesis,
are related exclusively to the eigenfunctions which fluctuate around
singular (exceptional) points.

\vspace{.6cm}

6. Concluding remarks.

Numerical investigations of concrete problems are seldom simpler
in the genuine
non-Hermitian formalism than those based on the Hermitian Hamiltonian    
$H_0$ and even on the effective Hamiltonian $H^{\rm eff}$.  
There are however many interesting problems 
of the many-body physics to which the standard
methods, including those based on  $H^{\rm eff}$,
cannot be applied. These are, e.g., questions related to the role of
EPs in many-channel problems,
or to the existence of other types of singular points. 
As a matter of principle, the
Hamiltonians $H_0$ or  $H^{\rm eff}$ can be used only 
for the description of many-particle
problems in which  internal degrees of freedom of the system 
(i.e. the eigenvalues) are excited.

Interesting questions are related to processes which 
occur without excitation of  internal degrees of freedom of the
system, i.e. without excitation of the eigenstates
of the non-Hermitian Hamiltonian. 
These questions can be studied only on the basis of 
the genuine Hamiltonian $\cal H$, in which 
not only the properties of the eigenstates but
also the fluctuations of the eigenfunctions 
around EPs are involved.
Photosynthesis is surely not the only example of such a process.   
These problems cannot be described
neither by $H_0$ nor  by $H^{\rm eff}$.
In literature, the high sensitivity of the eigenfunctions of the 
non-Hermitian Hamilton operator $\cal H$ and its
relation to the dynamics of open quantum systems is almost not 
studied up to now.

Another interesting result is that reordering processes under the
influence of EPs occur also in optics. The Dicke phenomenon of
superradiance is well known for many years. Only recently, the
corresponding long-living
subradiant  modes could be identified experimentally
\cite{kaiser}.

A further open question in non-Hermitian quantum physics is
the existence of 
singularities which are different from the EPs.
Their mathematical and physical
meaning should be clarified. An example are the points 
at which the  eigenfunctions $\Phi_i$
of the non-Hermitian Hamilton operator 
are orthogonal (and not biorthogonal).

Summarizing, it can be stated that the 
{\it attractiveness of non-Hermitian
quantum physics for the description of open quantum systems}
consists, above all, in the fact that
some basic problems of the standard Hermitian quantum physics
of closed systems do not appear
in the non-Hermitian quantum physics of open systems.
One of many examples is that the "Schr\"odinger cat" is not shielded
in an open quantum system from a direct observation.
More examples can be found in \cite{pra95}.
Studying concrete physical systems, it has 
been possible to explain 
some ``puzzling'' experimental results 
in the framework of the
non-Hermitian quantum physics. Other results will surely provide 
the basis for further interesting studies, including applications.


\vspace{1cm}



\begin{thebibliography}{99}

\bibitem{comment1}
We underline that we consider open quantum
systems. This should not be confused with the 
consideration of  PT-symmetric systems
the nature of which is different,  as stated 
in, e.g.,  C.M. Bender, Journal of Physics: 
Conference Series {\bf 631}, 012002 (2015).

\bibitem{feshbach} H. Feshbach, Ann. Phys. (NY) {\bf 5}, 357 (1958)\\
 H. Feshbach, Ann. Phys. (NY) {\bf 19}, 287 (1962)\\
P.O. L\"owdin, J. Math. Phys. {\bf 3}, 969 (1962)

\bibitem{stockmann} 
H.J. St\"ockmann,
{\it	Quantum Chaos. An Introduction}, Cambridge University Press 1999

\bibitem{top} 
 I. Rotter, J. Phys. A {\bf 42}, 153001 (2009)

\bibitem{sergi} A. Sergi and P.V. Giaquinta, Entropy {\bf 18}, 451
  (2016)

\bibitem{ro91} 
 I. Rotter, Rep. Prog. Phys. {\bf 54}, 635 (1991)

\bibitem{ropp} 
I. Rotter and J.P. Bird, 
Rep. Prog. Phys. {\bf 78}, 114001 (2015)

\bibitem{kato} T. Kato, 
{\it Perturbation Theory for Linear Operators}, Springer,
Berlin 1966

\bibitem{mois} 
 N. Moiseyev, {\it Non-Hermitian Quantum Mechanics}, 
 Cambridge University Press 2011

\bibitem{ep} 
J. Doppler, A. A. Mailybaev, J. B\"ohm,	U. Kuhl, A. Girschik,
F. Libisch, T. J. Milburn, P. Rabl,	N. Moiseyev, and S. Rotter,
Nature {\bf 537}, 76 (2016)\\
 H. Xu,	D. Mason, L. Jiang, and J. G. E. Harris,
Nature {\bf 537}, 80 (2016)\\
T.E. Lee, Phys. Rev. Lett. {\bf 116}, 133903 (2016)

\bibitem{pra95} 
H. Eleuch and I. Rotter, Phys. Rev. A {\bf 95},
022117 (2017)

\bibitem{muro} 
 M. M\"uller and I. Rotter, Phys. Rev. A {\bf 80}, 042705 (2009)

\bibitem{lapses} 
A. Yacoby, M. Heiblum, D. Mahalu, and H. Shtrikman,
Phys. Rev. Lett. {\bf 74}, 4047 (1995)\\
R. Schuster, E. Buks, M. Heiblum, D. Mahalu, V. Umansky,
and H. Shtrikman, Nature (London) {\bf 385}, 417 (1997)\\
M. Avinun-Kalish, M. Heiblum, O. Zarchin, D. Mahalu, and
V. Umansky, Nature (London) {\bf 436}, 529 (2005)

\bibitem{savin} 
Y.V. Fyodorov and D.V. Savin, Phys. Rev. Lett. {\bf 108}, 184101 (2012)\\
J.B. Gros, U. Kuhl, O. Legrand, F. Mortessagne, E. Richalot, and
D.V. Savin,  Phys. Rev. Lett. {\bf 113}, 224101 (2014)

\bibitem{comment2} 
In contrast to the definition that is used in, 
for example, nuclear physics, we define the complex energies 
before and after diagonalization of the non-Hermitian
Hamiltonian by 
$\varepsilon_k = e_k + \frac{i}{2} \gamma_k$ and  
${\cal E}_k = E_k + \frac{i}{2} \Gamma_k$, respectively, 
with $\gamma_k \le 0$ and $\Gamma_k \le 0$ for decaying states. 
This definition will be useful when discussing systems with {\it 
gain} (positive widths) and {\it loss} (negative widths), see
e.g. \cite{pre95}. 

\bibitem{pre95} 
H. Eleuch and I. Rotter, Phys. Rev. E {\bf 95},
062109 (2017)

\bibitem{engel} 
G. S. Engel, T. R. Calhoun, E. L. Read, T.-K. Ahn, T. Mancal,
Y.-C. Cheng, R. E. Blankenship, and G. R. Fleming, 
Nature {\bf 446}, 782 (2007)\\
H. Lee, Y.-C. Cheng, and G. R. Fleming, Science {\bf 316}, 1462 (2007)

\bibitem{engel-ed}
M. Mohseni, Y. Omar, G.S. Engel, and M.B. Plenio (eds.), {\it Quantum
Effects in Biology}, Cambridge University Press, Cambridge, UK, 2014  

\bibitem{fleming}  
H. Dong and   G. R. Fleming,
J. Phys. Chem. B {\bf 118}, 8956 (2014)

\bibitem{romero} 
E. Romero, R. Augulis,	 V.I. Novoderezhkin,	 M. Ferretti,	
J. Thieme, D. Zigmantas, and R. van Grondelle,	
Nature Physics {\bf 10}, 676 (2014)

\bibitem{kaiser}
W. Guerin, M.O. Araujo, and R. Kaiser, 
Phys. Rev. Lett. {\bf 116}, 083601 (2016) 

\end{thebibliography}
\end{document}